\documentclass[11pt]{article}
\newsavebox{\PSLASH}
\sbox{\PSLASH}{$p$\hspace{-1.8mm}/}

\begin{document}
\title{\large \bf  Revisiting effective Einstein equations on a 3-brane in the presence of torsion}
\author{ S. Khakshournia$^{1}$
	\footnote{Email Address: skhakshour@aeoi.org.ir} and R.
	Mansouri$^{2,3}$ \footnote{Email Address: mansouri@ipm.ir}\\
	\\
	$^{1}$Nuclear Science and Technology Research Institute (NSTRI), Tehran, Iran\\
	$^{2}$Department of Physics, Sharif University of
	Technology, Tehran, Iran\\
	$^{3}$School of Astronomy, Institute for Research in Fundamental
	Sciences (IPM), Tehran, Iran \\
}\maketitle
\[\]
\[\]

 \begin{abstract}

 	The effective Einstein equations on a 3-brane embedded in a 5-dimensional Riemann-Cartan bulk spacetime are revisited. Addressing the shortcomings in the hitherto published junction conditions on the brane in the presence of torsion, we have elaborated on our general form of the junction conditions recently published. Applying our general junction conditions, we have formulated the effective Einstein equations on a $ Z_{2} $ symmetric brane in a standard form highlighting the difference to those published so far.\\

 \noindent 
 Keywords: 3-Brane world, Torsion, Junction conditions, effective Einstein equations\\
 \end{abstract}
 \hspace{1.5cm}
 \newpage
 \section{Introduction}

 There has been considerable interest over two decades in the brane-world scenarios, motivated by the string theory. Therein, our universe is a 3-brane embedded in a 5-dimensional bulk spacetime. Although the fifth dimension orthogonal to the brane is noncompact, gravitation is localized on the brane reproducing effectively four-dimensional gravity at large distances due to the warped geometry of the spacetime \cite{Randall99} (see also \cite{Maartens010} for a review).\\ In this context, the effective field equations on a 3-brane embedded in a five-dimensional bulk spacetime were derived via an elegant geometrical projection approach by Shiromizu et al. \cite{Shiromizu2000}, using the Gauss-Codazzi equations, junction conditions for the extrinsic curvature on the brane, and $ Z_{2} $ symmetry. These equations involve the higher-dimensional corrections to the ordinary Einstein equations, which are determined by a term quadratic in the energy-momentum tensor of the brane, as well as a curvature term from the bulk spacetime.\\
 What if the torsion is not vanishing in the braneworld model? The junction conditions, being crucial for this case, are not well established yet. Even a standard form of the projected Einstein equations on the brane is not yet at hand, as can be seen in the different results of  \cite{da2009braneworld, Maier2012}. The differences are mainly due to different assumptions on the junction conditions of the torsion tensor across the brane, and also due to the form of the torsion tensor assumed taking into account its antisymmetric properties. Hoff da Silva et al. \cite{da2009braneworld} (see also \cite{da2010possible}) assumed a form for the jump of torsion tensor across the brane not reflecting the required antisymmetric properties, resulting in no torsion contribution to the jump of brane extrinsic curvature tensor, and some unwanted orthogonal components of the brane energy-momentum tensor. Then, based on it, they obtained the projected effective Einstein equations in a 4-dimensional arbitrary manifold embedded in a 5-dimensional Riemann-Cartan manifold, different from \cite{Maier2012}.\\
 Maier et al. \cite{Maier2012} (see also \cite{Maier2009}) assumed the torsion in the bulk to be continuous across the brane allowing just the first derivatives appearing in the effective field equations on the 3-brane to be discontinuous, similar to the familiar assumptions on the metric. We know however, that the corresponding assumptions for the metric in Riemannian spacetimes can be made without loss of generality due to the freedom of the choice of the coordinates, which leads to the so-called admissible coordinates. Once we have used the freedom of the coordinates to make the metric continuous across the brane, there remains no more freedom to make all the components of the torsion continuous across the brane. Therefore, this assumption which is assumed to be allowed in general in the paper \cite{Maier2012} is not allowed except for very special cases such as considered at the end of that paper; a fact we have elaborated on it in this paper. The example discussed in \cite{Maier2012} is a very special vacuum bulk where the torsion is assumed to be highly reduced to just one component and is trivially continuous across the brane. Hence, this case is very singular, and one does not learn much about brane cosmology with torsion in general. In addition, the authors in \cite{Maier2012} presupposed the brane energy-momentum tensor in the presence of torsion to be symmetric which is too restrictive and not correct in general, and to be tangent to the brane, which is unnecessary.\\
 Having noted these shortcomings in the junction conditions across the brane in the presence of torsion, our aim in this paper is to revisit the effective Einstein equations on a 3-brane with non-zero torsion while using our recently presented recipe for the junction conditions on the brane \cite{khakshournia2020}. We, therefore, claim to use the most general formulation for the junction conditions in the presence of torsion to obtain the effective Einstein equations on the brane in a five-dimensional brane cosmology.\\ 
 The paper is organized as follows. In section 2, we briefly review basic aspects of the Riemann-Cartan spacetime and give all equations we need in this work. In section 3, we summarize our formulation for junction conditions in the presence of torsion, addressing the shortcomings in junction conditions so far in the literature. Section 4 is devoted to a standard formulation of the effective Einstein equations on a $Z_2$ symmetric brane based on our junction conditions presented in section 3, noting some of the main features induced from the bulk. We also elaborate on some delicate deviations of our standard form of the effective Einstein equations from that published so far in the literature.  \\
 
 \textit{Conventions.}  We use square brackets [F] to denote the jump of any quantity F across the brane. Throughout this manuscript, tildas will indicate Riemannian variables related to the Christoffel symbols only. We also adopt a 5-dimensional spacetime metric with signature ( -, +, +, +,+). Capital Latin indices are used for the coordinates in the five-dimensional manifold, whereas small Latin indices for $ n $-dimensional manifold, and Greek indices denotes the four-dimensional brane coordinates. We have adopted the definitions and the conventions of \cite{Kranas2019, bressange2000extension} for torsion.\\

  \section{Riemann-Cartan manifold}
 In an  $ n $-dimensional Riemann-Cartan spacetime manifold endowed with a metric tensor $ {}^{(n)}g_{_{ab}}$, having an asymmetric affine connection ${}^{(n)}{\Gamma^{a}}_{bc} $, the torsion tensor is defined by the antisymmetric component of the affine connection as
  \begin{equation}\label{tortdef}
   		{}^{(n)}{T^{a}}_{bc}={}^{(n)}{\Gamma^{a}}_{cb}-{}^{(n)}{\Gamma^{a}}_{bc},
  \end{equation}
   ensuring that the torsion tensor	$ {}^{(n)}{T^{a}}_{bc} $ is antisymmetric in its second and third indices. Demanding that the metric tensor is covariantly constant, i.e. ${}^{(n)} \nabla_{c}g_{ab}=0 $, the following decomposition of the asymmetric connection can be done:
   \begin{equation}\label{tordecom}
   		{}^{(n)}{\Gamma^{a}}_{bc}= 	{}^{(n)}\tilde{{\Gamma}}^{a}_{\;\,bc}
   	+	{}^{(n)}{K^{a}}_{bc},
   \end{equation}
 where $ {}^{(n)}\tilde{{\Gamma}}^{a}_{\;\,bc} $ denotes the usual Levi-Civita connection given by 
 \begin{equation}\label{tordecom}
    	 \tilde{{\Gamma}}^{a}_{\;\,bc}=\frac{1}{2}g^{ad}(\partial_{b}g_{cd}+\partial_{c}g_{bd}-\partial_{d}g_{bc}),
    \end{equation}
 and $ {}^{(n)}{K^{a}}_{bc} $ are the components of the contorsion tensor
 of the connection given in terms of the components of the torsion by
   \begin{equation}\label{torcotdef}
   		{}^{(n)}K_{abc}=\frac{1}{2}({}^{(n)}T_{bac}
   	+{}^{(n)}T_{cab} -{}^{(n)}T_{abc}),
   \end{equation}
  with ${}^{(n)} K_{abc}=-{}^{(n)}K_{bac} $.
   In the presence of torsion, the components of the curvature tensor have the usual expression
   \begin{equation}\label{torriedef}
      R^{a}_{\;\,bcd}=\partial_{c}\Gamma^{a}_{\;\,bd}-\partial_{d}\Gamma^{a}_{\;\,bc}+\Gamma^{a}_{\;\,ec}\Gamma^{e}_{\;\,bd}-\Gamma^{a}_{\;\,ed}\Gamma^{e}_{\;\,bc}.
    \end{equation} 
   The remaining symmetries of the curvature tensor  are $ R_{abcd}=R_{[ab][cd]}$. In the presence of torsion, the Bianchi identities acquire a non-zero right-hand side comparing to its Riemannian analogue as \cite{Kranas2019}
     \begin{equation}\label{torbiaid1}
      {}^{(n)}R^{a}_{\;\,[bcd]}={}^{(n)}\nabla_{[b}{}^{(n)}T^{a}_{\;\,cd]}+{}^{(n)}T^{a}_{\;\,e[b}{}^{(n)}T^{e}_{\;\,cd]}.
      \end{equation}
      \begin{equation}\label{torbiaid2}
      	{}^{(n)}\nabla_{f}{}^{(n)} R^{a}_{\;\,bcd}=-{}^{(n)}R^{a}_{\;\,be[f}{}^{(n)}T^{e}_{\;\,cd]}.
      \end{equation}
   Contraction of the identity (\ref{torbiaid1}) twice yields
  \begin{equation}\label{torcyccont}
      2{}^{(n)}G_{[ab]}={}^{(n)}\nabla_{c}{}^{(n)}T^{c}_{\;\,ab}+{}^{(n)}\nabla_{a}{}^{(n)}T^{c}_{\;\,bc}-{}^{(n)}\nabla_{b}{}^{(n)}T^{c}_{\;\,ac}
      +{}^{(n)}T^{c}_{\;\,dc}{}^{(n)}T^{d}_{\;\,ba},
   \end{equation}
   indicating that the Einstein tensor $ {}^{(n)}G_{ab} $ is asymmetric, and hence, in general, the energy-momentum tensor of spacetime  $ {}^{(n)}T_{ab} $, is asymmetric as well.
   The decomposition of the curvature tensor into its traces and a trace-free component can still be done with noticing that in the presence of torsion the Ricci tensor is asymmetric: 
    \begin{equation}\label{wyeldecomp}
   {}^{(n)}R_{abcd} = {}^{(n)}C_{abcd} + \frac{2}{n-2}({}^{(n)}R_{a[c}{}^{(n)}g_{d]b}-{}^{(n)}R_{b[c}{}^{(n)}g_{d]a})-\frac{2}{(n-1)(n-2)}{}^{(n)}R{}^{(n)}g_{a[c}{}^{(n)}g_{d]b},
\end{equation}
 where the trace-free component $ {}^{(n)}C_{abcd} $ is referred to as the Weyl-Cartan curvature tensor with the remaining symmetries as $ C_{abcd}=C_{[ab][cd]}$. \\
    The curvature tensor (\ref{torriedef}) can be expressed through the Riemann tensor (curvature tensor
     depending only on the metric), covariant derivative (torsionless) and contorsion as \cite{Kranas2019}  
   \begin{equation}\label{curvtdecom}
           {}^{(n)}R^{a}_{\;\,bcd}=	{}^{(n)}\tilde{R}^{a}_{\;\,bcd}+ {}^{(n)} Q^{a}_{\;\,bcd},
    \end{equation}
    where
    \begin{equation}\label{torcurv1}
          	{}^{(n)}Q^{a}_{\;\,bcd}= {}^{(n)}\nabla_{c}	{}^{(n)}K^{a}_{\;\,bd}-	{}^{(n)}\nabla_{d}	{}^{(n)}K^{a}_{\;\,bc}+	{}^{(n)}K^{a}_{\;\,ec}	{}^{(n)}K^{e}_{\;\,bd}-	{}^{(n)}K^{a}_{\;\,ed}	{}^{(n)}K^{e}_{\;\,bc}.
    \end{equation}
       Similar expressions can be written for the Ricci tensor $  {}^{(n)}R_{ab}={}^{(n)}\tilde{R}_{ab}+{}^{(n)}Q_{ab} $ with
    \begin{equation}\label{torcurv2}
         	{}^{(n)}Q_{ab}= {}^{(n)}\nabla_{c}	{}^{(n)}K^{c}_{\;\,ab}-	{}^{(n)}\nabla_{b}	{}^{(n)}K^{c}_{\;\,ac}+	{}^{(n)}K^{c}_{\;\,dc}	{}^{(n)}K^{d}_{\;\,ab}-	{}^{(n)}K^{c}_{\;\,ad}	{}^{(n)}K^{d}_{\;\,cb},
      \end{equation}
      and for the Ricci scalar
      \begin{equation}\label{torricci}
       {}^{(n)}R={}^{(n)}\tilde{R}+{}^{(n)}Q 
       \end{equation}
         with
      \begin{equation}\label{torcurv3}
           	{}^{(n)}Q=-2{}^{(n)}\nabla_{e}{}^{(n)}K^{ce}_{\;\,c}-	{}^{(n)}K_{cd}^{\;\,d}{}^{(n)}K^{ce}_{\;\,e}+{}^{(n)}K_{cde}{}^{(n)}K^{ced}.
     \end{equation}
     The 5-dimensional action for the gravitational field and matter in the Einstein-Cartan theory is written as \cite{Poplawski2014}
     \begin{equation}\label{toraction}
     S_{EC}=\int(\frac{1}{2\kappa_{5}^2} 	{}^{(5)}R\sqrt{- 	{}^{(5)}g}+{}^{(5)}L_{m})d^{5}x,
     \end{equation}
      where the curvature scalar $ {}^{(5)}R $ is given by (\ref{torricci}),  $ {}^{(5)}g $ is the determinant of the metric tensor $ {}^{(5)}g_{AB} $, $ \kappa_{5}^2 $ is the 5-dimensional gravitational coupling constant, and $ {}^{(5)}L_{m} $ is the Lagrangian density for matter.
     The stationarity of action (\ref{toraction}) with respect to the variation of the metric tensor gives the Einstein equations
      \begin{equation}\label{torfieldeq1}
      G_{AB}=\kappa_{5}^2 T_{AB},
      \end{equation}
     The stationarity of action (\ref{toraction}) with respect to the variation of the torsion tensor yields the Cartan equations
      \begin{equation}\label{torfieldeq2}
    {T^{A}}_{BC}+ \delta^{A}_{B} {T^{D}}_{CD}- \delta^{A}_{C} {T^{D}}_{BD}=\kappa_{5}^2 {S^{A}}_{BC},
     \end{equation}
     where  $ {S^{A}}_{BC}=2/(-	{}^{(5)}g)^{1/2} \delta {}^{(5)}L_{m}/\delta {K^{BC}}_{A}$ is the spin tensor representing the density of the intrinsic angular momentum in the matter. Since the Cartan equations (\ref{torfieldeq2}) are linear and algebraic, the torsion tensor vanishes outside material bodies where the
      spin density is zero.
      \section{Junction conditions}
      
      In this section, we review the relevant results concerning the junction conditions with torsion as obtained in \cite{khakshournia2020}. Consider a $ 3 $-brane, a (3 + 1)-dimensional timelike hypersurface, embedded in a $ 5 $-dimensional bulk spacetime.
      Let $X^{A}$ be an admissible coordinate system in a coordinate neighborhood that includes the brane extending into both Riemann-Cartan spacetimes $\cal M^{\pm}$. The parametric equation of the brane is written as $\Phi (X^{A})=0$, where $\Phi$ is a smooth function. The domains in which $\Phi$ is positive or negative are contained in $\cal M^{+}$ or $\cal M^{-}$, respectively. The metric across the two domains can be written as the distribution-valued tensor
        \begin{equation}\label{distmetric}
          g_{AB}=	g^{+}_{AB}\Theta(\Phi)+	g^{-}_{AB}\Theta(-\Phi),
        \end{equation}
      where $ \Theta(\Phi) $ is the step function and $ g^{-}_{AB} $ and $ g^{+}_{AB} $ are metrics in $ \cal M^{-} $ and $ \cal M^{+} $, continuously 
      joined on $ \Sigma $. The corresponding tangent vectors on the brane are $ e_{(\mu)}=\partial/\partial x^{\mu} $, and a normal vector is defined 
      by $n_{A}=|\alpha|^{-1}\partial_{A}\Phi$, where $\alpha$ is a normalizing factor such that $  n_{A}n^{A}=1$,  The relevant jumps on the brane expressed in the admissible coordinates $X^{A}$ must vanish:
       $[g_{AB}]=[n^{A}]=[e^{A}_{~\mu}]=[\alpha]=0$.  Given the induced metric $  g_{\mu\nu}=e_{(\mu)}.e_{(\nu)} $ on the brane, the completeness relations for the basis are written as
       \begin{equation}\label{completerel}
        g^{AB}= g^{\mu\nu}e^{A}_{~\mu}e^{B}_{~\nu}+ n^{A}n^{B},
       \end{equation}
       where $ g^{\mu\nu} $ is the inverse of the induced metric. 
       The metric continuity condition implies that the tangential derivatives of the metric are continuous $[g_{AB,C}]e^{C}_{~\mu}=0$. 
        However, the normal derivative of the metric, $g_{AB,C}n^{C}$, may be discontinuous, leading to
       \begin{equation}\label{metricdisc}
          [g_{AB,C}] =\gamma_{AB}n_{C},	 
        \end{equation}
        where the tensor field $\gamma_{AB}$ is given by $ \gamma_{AB}= [g_{AB,C}]n^{C} $.
        Similarly, we assume the torsion to be continuous along the brane. Technically it means that the purely tangential part of the torsion tensor is continuous across the brane  \cite{bressange2000extension}:
        \begin{equation}\label{tortancomt}
               	e^{A}_{~\mu}e^{B}_{~\nu}e^{C}_{~\sigma}[T_{ABC}]=[T_{\mu\nu\sigma}]=0. 
        \end{equation}
       In general, we have to assume that the transverse components of the torsion are discontinuous, as there is no more freedom in the choice of the coordinates to make these transverse components continuous. The assumption of the continuity of transverse components of the torsion in \cite{Maier2012} is a severe restriction to the generality of the junction conditions and for the effective Einstein equations on the brane. Now, let us assume the general case of the continuity of the bulk torsion tensor along the brane as expressed in (\ref{tortancomt}), and the discontinuity of the bulk torsion tensor across the brane as
          \begin{equation}\label{tordisc}      
      [{T^{A}}_{BC}]=\zeta_{BC}n^{A},
          \end{equation}
     where the antisymmetric tensor $ \zeta_{BC} $, being consistent with the antisymmetric property of ${ T^{A}}_{BC} $ on its last two indices, is given by $ \zeta_{BC}=n_{A}[{T^{A}}_{BC}] $. We will see in the following how suitable this natural choice to express the discontinuity of the torsion tensor is, in contrast to the choice of \cite{da2009braneworld} neglecting the trivial symmetry feature of the torsion.\\
     To generate the present discontinuity of the torsion, we first substitute (\ref{tordisc}) into the Cartan equation (\ref{torfieldeq2}) to obtain
     \begin{equation}\label{cartancont}
     \zeta_{BC}n^{A}+ \delta^{A}_{B} \zeta_{CD}n^{D}-\delta^{A}_{C} \zeta_{BD}n^{D}=\kappa_{5}^2 [{S^{A}}_{BC}].
     \end{equation}
     Noticing the antisymmetric property of the spin tensor on the last two indices, the contraction of (\ref{cartancont}) with the normal vector on each of the antisymmetric indices leads to
     \begin{equation}\label{spinjump1}
     n^{B}[{S^{A}}_{BC}]=0,\hspace{1cm} n^{C}[{S^{A}}_{BC}]=0.
     \end{equation}
     Now, the contraction with $ n_{A} $ yields
     \begin{equation}\label{spinjump2}
    \kappa_{5}^2 n_{A}[{S^{A}}_{BC}]=\zeta_{BC}+\zeta_{CD}n^{D}n_{B}-\zeta_{BD}n^{D}n_{C},
     \end{equation}
     indicating the orthogonal components of the spin tensor (\ref{spinjump1}), constructed out of contracting with each of its antisymmetric indices, to be 
     continuous across the brane, while the normal components $ n_{A}[{S^{A}}_{BC}] $ are to be discontinuous as seen from (\ref{spinjump2}).
    A further contraction of (\ref{spinjump2}) with the normal vector yields
     \begin{equation}\label{spinjump3}
     n_{A}n^{B}[{S^{A}}_{BC}]=n_{A}n^{C}[{S^{A}}_{BC}]=0,
     \end{equation}
     showing the expected requirement on the spin tensor to generate the discontinuity (\ref{tordisc}).\\
    Now, based on the discontinuities (\ref{metricdisc}) and (\ref{tordisc}) across the brane, we obtain the following expression for the brane energy-momentum tensor $S_{AB} $ \cite{khakshournia2020}:
        \begin{eqnarray}\label{surfaceengten}
           \kappa^{2}_{5}  S_{AB}&=&-\frac{1}{2}(\gamma n_{A}n_{B}+g_{AB}\gamma_{CD}n^{C}n^{D}-\gamma_{BD}n^D n_A-\gamma_{AD}n^{D}n_{B}+\gamma_{AB}-\gamma g_{AB})\\ \nonumber
           	  &+& \frac{1}{2}(\zeta_{AD} n_B n^{D}-\zeta_{BD} n_A n^{D}-\zeta_{AB}),	
         \end{eqnarray}
       where $ \kappa^{2}_{5} = 8\pi G_{5}$, with $ G_{5} $ being the five-dimensional Newton's constant. Note that, due to the contribution of the jump of the torsion, this energy-momentum tensor is non-symmetric except in the very special case of the torsion being continuous across the brane (as assumed in \cite{Maier2012}). We have already seen that this is a very special case not to be used in general In addition, the asymmetric tensor $ S_{AB} $ is automatically purely tangent to the brane, i.e. the conditions $ S_{AB}n^{A}=0 $ and  $S_{AB}n^{B}=0 $ hold simultaneously, as expected. Hence, there is no need for an extra assumption of vanishing the transverse components of the brane energy-momentum tensor as in \cite{Maier2012} , or to impose (\ref{spinjump3}) as an extra condition in order for the brane energy-momentum tensor to be tangent to brane as done in \cite{bressange2000extension}, or to force any emergent transverse components to vanish due to an unfortunate choice of the torsion tensor in \cite{da2009braneworld}.\\
       On the other hand, the Cartan equation (\ref{torfieldeq2}) can be rewritten as
        \begin{equation}\label{torfieldeq22}
           {T^{A}}_{BC}= \kappa_{5}^2({S^{A}}_{BC}+\frac{1}{3}\delta^{A}_{B} {S^{D}}_{CD}- \frac{1}{3}\delta^{A}_{C} {S^{D}}_{BD}).
            \end{equation}
       It shows that in general, the jump of the torsion tensor across the brane is a function of the spin tensor; it vanishes only if the spin tensor itself is zero. The unfortunate assumption of the vanishing the jump of torsion across the brane in \cite{Maier2012} is, therefore, a severe restriction of the general case and making their Einstein equations not valid in general. The example discussed by the authors at the end of their paper, however, assuming a vacuum configuration in the bulk, does agree with the special conditions already discussed. \\ 
        Defining the extrinsic curvature tensor of the brane 
        \begin{equation}\label{extrintendef}
                 {\cal K}_{\mu\nu}=e^{A}_{~\mu}e^{B}_{~\nu}\nabla_{A}n_B,
         \end{equation}
      where the covariant derivative in the definition (\ref{extrintendef}) is made with torsion. The jump of the extrinsic curvature across the brane is calculated to be
         \begin{equation}\label{extrinjump}
         [{\cal K}_{\mu\nu}]=\frac{1}{2}(\gamma_{\mu\nu}+\zeta_{\mu\nu}).
        \end{equation}
       Eq. (\ref{extrinjump}) may be rewritten as
     \begin{equation}\label{extrintenexpcom}
          [{\cal K}_{\mu\nu}]=\frac{1}{2}[g_{AB,C}]n^{C}e^{A}_{~\mu}e^{B}_{~\nu}+\frac{1}{2}[T_{ABC}]n^{A}e^{B}_{~\mu}e^{C}_{~\nu}, 
     \end{equation}
       where (\ref{metricdisc}) and (\ref{tordisc}) have been used. It is now explicitly seen from this form of the junction conditions that a torsion contribution to the brane extrinsic curvature is present only if there is a jump in the bulk torsion tensor across the brane.  Making the decomposition $ S_{\mu\nu}=S_{AB}e^{A}_{~\mu}e^{B}_{~\nu} $ from (\ref{surfaceengten}) and using $ (\ref{extrinjump}) $, the final form of junction conditions in the presence of torsion may then be written as
     
     \begin{equation}\label{ijct}
        [{\cal K}_{\mu\nu}]-[{\cal K}]g_{\mu\nu}=-\kappa^{2}_{5} S_{\mu\nu},
      \end{equation}
     or equivalently

   \begin{equation}\label{ijct-2}
        [{\cal K}_{\mu\nu}] = -\kappa^{2}_{5} (S_{\mu\nu} -\frac{1}{3}S g_{\mu\nu}),
      \end{equation}
   where $ [{\cal K}]=\gamma $. Therefore, the junction conditions seem formally the same as in general relativity. The difference is implicit in the non-symmetric nature of $ [{\cal K}_{\mu\nu}] $ being decomposed into a Riemann part $ \gamma_{\mu\nu} $ and a Cartan part $ \zeta_{\mu\nu} $,  shown in  (\ref{extrinjump}). This result is in clear contrast with that obtained in \cite{da2009braneworld}, expressing that the Israel junction conditions unexpectedly remain the same as in the general relativity. This fault is due to the unfortunate choice of the jump of the torsion tensor leading to an expression for the brane extrinsic curvature tensor which is symmetric with no torsion contribution included.\\
  The results of this section may be itemized as follows:
      \begin{itemize}
  \item The bulk metric can be made continuous across the brane. However, its normal derivative  is  discontinuous, and given by Eq.( \ref{metricdisc}).
  \item In general, once we have assumed the continuity of the bulk metric tensor across the brane, there is not enough freedom left to make the bulk torsion continuous as well. Therefore,  we have to assume the bulk torsion has a jump across the brane as singled out by Eq. (\ref{tordisc}).
  \item The brane extrinsic curvature is not symmetric in the presence of torsion. The jump of it across the brane is naturally decomposed into a symmetrical part $ \gamma_{\mu\nu} $ and an antisymmetrical part $ \zeta_{\mu\nu} $ as seen in Eq. (\ref{extrinjump}).
  \item In general, the brane energy-momentum tensor is also asymmetric, as can be seen from Eq. (\ref{surfaceengten}).
  \item The junction conditions are formally the same as in general relativity, except the asymmetric nature of the tensors involved.
  \item In case the torsion has no jump across the brane, it is seen from Eqs. (\ref{surfaceengten}) and (\ref{extrinjump}) that both the brane extrinsic curvature tensor and the brane energy-momentum tensor are symmetric. This is a very special case studied in \cite{Maier2012}.
      \end{itemize} 

 \textit{\textbf{The case of $Z_{2}$ symmetry}} \\
    The most general case we studied above can be reduced to the more simple case of the spacetime with the $ Z_{2} $ symmetry on the brane at the fixed point. In this case, taking into account $ {\cal K}^{+}_{\mu\nu}=-{\cal K}^{-}_{\mu\nu}$ due to the $ Z_{2} $ symmetry and the junction conditions (\ref{ijct-2}), the generalized extrinsic curvature of the brane which includes the torsion contribution, can be written uniquely in terms of the asymmetric energy-momentum tensor of the brane:
     \begin{equation}\label{extrintenexp}
     	{\cal K}^{+}_{\mu\nu}=-{\cal K}^{-}_{\mu\nu}=-\frac{1}{2}\kappa^{2}_{5}\left(S_{\mu\nu}-\frac{1}{3}Sg_{\mu\nu}\right). 
     \end{equation}
  Hence, using $ {\cal K}^{+}_{\mu\nu}=-{\cal K}^{-}_{\mu\nu} $, $ g^+_{AB,C}=-g^-_{AB,C} $, and $ T^{+}_{ABC}=-T^{-}_{ABC} $, from 
  (\ref{extrintenexpcom}) we obtain 
       
     \begin{equation}\label{extrintenexpcomz2}
            {\cal K}_{\mu\nu}=\frac{1}{2}g_{AB,C}n^{C}e^{A}_{~\mu}e^{B}_{~\nu}+\frac{1}{2}T_{ABC}n^{A}e^{B}_{~\mu}e^{C}_{~\nu}. 
     \end{equation}
  Authors in \cite{Maier2012}, using a "pill-box" integration of Einstein equations in the presence of torsion through the brane, obtain a similar result to our equation (\ref{ijct}) or (\ref{ijct-2}) expressed in their Eq. (26). They, however, make the assumption that the brane energy-momentum tensor on the right-hand side of this equation is symmetric, in contrast to the fact that the brane extrinsic curvature in the left-hand side of their Eq. (26) is built with the asymmetric connection and is also asymmetric. This has led to their equation (28) for an asymmetric extrinsic curvature tensor which is inconsistent with their Eq. (26). However, it is worth mentioning that for highly symmetric spacetimes the energy-momentum tensor may retain its familiar symmetry and assuming the very specific case of a continuous torsion, as is the case in the concrete example at the end of the paper \cite{Maier2012}, this fallacy does not appear.

     \section{Effective equations on the brane}
     
     We are now prepared to drive the effective Einstein equations on the brane. Consider a Gaussian coordinate system as 
     \begin{equation}
            ds^{2}=dy^{2}+g_{\mu\nu}dx^{\mu}dx^{\nu},
              \label{Edef}
      \end{equation}              	
    where $ y $ denotes the extra dimension orthogonal to the brane being at $ y = 0 $, without loss of generality. As a condition on the coordinate in the direction of the extra dimension, we may write  $ n_{A}dX^{A} = dy $, with $ n^{A} $ the spacelike unit normal to
    the $Z_2$ symmetric brane to be able to compare the results with those already published.  Following the approach adopted in \cite{Shiromizu2000}, we start with the Gauss equation   
     \begin{equation}\label{Gausseq}
     R_{\sigma\mu\rho\nu}
     	= R_{ABCD}
     	e^A_{~\sigma} e^B_{~\mu}e^C_{~\rho} e^D_{~\nu}
     	 +{\cal K}_{\mu\nu}{\cal K}_{\sigma\rho}
     	-{\cal K}_{\mu\rho}{\cal K}_{\sigma\nu}, 
     \end{equation}
     and the Codazzi equation in the presence of torsion \cite{Maier2012}
     \begin{equation}\label{Codacci}
            	 \nabla_\mu {\cal K}^{\mu}_{\;\,\nu} -  \nabla_\nu {\cal K}
            	= R_{AB}{n^A} e^B_{~\nu}-T^{\sigma}_{~\mu\nu}\cal K^{\mu}_{~\sigma} \,.
      \end{equation} 
      Contracting the Gauss equation (\ref{Gausseq}) finally yields
         \begin{eqnarray} \label{4dee1}
          	G_{\mu\nu}
              &	=&\left(R_{BD}-\frac{1}{2}g_{BD}R\right) e^B_{~\mu}e^D_{~\nu}-R_{ABCD}
             n^A e^B_{~\mu}n^C e^D_{~\nu}+q_{\mu\nu}R_{BD} n^Bn^D\nonumber \\
             &+&{\cal K}_{\mu\nu}{\cal K}
           	-{\cal K}_{\mu\sigma}{\cal K}^{\sigma}_{~\nu}
          	-\frac{1}{2}g_{\mu\nu}({\cal K}^{2}
          	-{\cal K}_{\mu\sigma}{\cal K}^{\sigma\nu}). 
          \end{eqnarray}                  
      Using 5-dimensional Einstein equations  
       \begin{equation}\label{5eequat}
        G_{AB}\equiv \tilde{G}_{AB}+L_{AB}=-\Lambda_5 g_{AB}+\kappa^{2}_{5}T_{AB},
       \end{equation}
         where $ \Lambda_{5} $ is the bulk cosmological constant, $ T_{AB} $ is the 5-dimensional energy-momentum tensor, and
         \begin{equation} 
        \tilde {G}_{AB}=\tilde{R}_{AB} -\frac{1}{2}\tilde Rg_{AB},
          \end{equation} 
          denotes the Riemannian part and  
          \begin{equation} 
            {L}_{AB}={Q}_{AB} -\frac{1}{2}Qg_{AB},
           \end{equation}
       represents the torsion part of the Einstein tensor with the torsion contributions $ {Q}_{AB} $ and $ Q $ given in (\ref{torcurv2}) and (\ref{torcurv3}), respectively.
       Using the decomposition of the 5-dimensional curvature tensor into its traces and a trace-free part named the Weyl-Cartan tensor as given in (\ref{wyeldecomp}), and the Einstein equations (\ref{5eequat}) to substitute the 5-dimensional energy-momentum tensor for the trace part of the curvature tensor, Eq. (\ref{4dee1}) takes the form
       \begin{eqnarray}	\label{4dee2}
     	G_{\mu\nu}
     	&=&-\frac{1}{2}\Lambda_{5} g_{\mu\nu} +{2 \kappa_5^2 \over 3}\left(T_{AB}
     	e^{A}_{~\mu} e^{B}_{~\nu}  
     	+\left(T_{AB}n^A n^B-{1 \over 4}T\right)
      g_{\mu\nu} \right)\nonumber\\
       &+& {\cal K} {\cal K}_{\mu\nu} -{\cal K}^{~\sigma}_{\mu}{\cal K}_{\nu\sigma} 
     	-{1 \over 2}g_{\mu\nu}
     	\left({\cal K}^2-{\cal K}^{\alpha\beta}{\cal K}_{\alpha\beta}\right)
      - E_{\mu\nu}, 
      \end{eqnarray}
     where
     \begin{equation}
     	E_{\mu\nu} \equiv C_{ABCD}n^A e_{~\mu}^{B}n^C e_{~\nu}^{D},
     	\label{Edef}
     \end{equation}
      is the electric part of the Weyl-Cartan tensor in the bulk induced on the brane. It is commonly assumed that the 5-dimensional energy-momentum tensor has the form
      \begin{equation}\label{decompengymom1}
     T_{AB}=\bar{T}_{AB}+S_{AB}\delta(y),
        \label{Edef}
       \end{equation}
       where $ \bar{T}_{AB} $ and $S_{AB} $ are the continuous and discontinuous components of the bulk energy-momentum, respectively.
       The discontinuous component $S_{AB} $, being total energy-momentum tensor of the brane, is decomposed into an ordinary matter part and a pure tension part as 
     \begin{equation}\label{decompengymom2}
       S_{\mu\nu}\equiv S_{AB}e^A_{~\mu} e^B_{~\nu}=\tau_{\mu\nu}-\sigma g_{\mu\nu}.
      \end{equation}
       Substituting Eqs. (\ref{decompengymom1}) and (\ref{decompengymom2}), as well as the expression (\ref{extrintenexp}) for the asymmetric extrinsic curvature of the brane into Eq. (\ref{4dee2}), we arrive at the final form of effective Einstein equations on the $ Z_{2} $ symmetric brane:
     \begin{eqnarray}	\label{4dee3}
        G_{\mu\nu}=-\Lambda_{4} g_{\mu\nu}+8\pi G_{N}\tau_{\mu\nu}+\kappa^{2}_{5}\Pi_{\mu\nu}-E_{\mu\nu}+F_{\mu\nu},
      \end{eqnarray}
      where the following definitions have been made:
     \begin{equation}
         \Lambda_{4}=\frac{1}{2}\Lambda_{5}+\frac{1}{12}\kappa_5^4\sigma^{2},
       	\label{Edef}
      \end{equation}
    \begin{equation}
         G_{N}=\frac{\kappa_5^4\sigma}{48\pi},
         	\label{Edef}
    \end{equation}
     \begin{equation}\label{Edef}
        \Pi_{\mu\nu}=-\frac{1}{4}\tau_{\mu\sigma}\tau_{\nu}^{~\sigma}+\frac{1}{12}\tau\tau_{\mu\nu}+\frac{1}{8}g_{\mu\nu}\tau_{\sigma\rho}\tau^{\sigma\rho}-\frac{1}{24}g_{\mu\nu}\tau^{2},
      \end{equation}
    \begin{equation}
       F_{\mu\nu}={2 \kappa_5^2 \over 3}\left( \bar {T}_{AB}
     	e^{A}_{~\mu}  e^{B}_{~\nu}+\left( \bar{T}_{AB}n^A n^B-{1 \over 4} \bar{T}\right)
       g_{\mu\nu} \right),          
       \label{fdef}
      \end{equation}
    where we have tried to use the same symbols as authors in \cite{da2009braneworld} and \cite{Maier2012}. From Eqs. (\ref{torcurv2}) and (\ref{torcurv3}), the (con)torsion terms in  $ G_{\mu\nu} $ can be visualized as 
        \begin{eqnarray}\label{fdett}
       G_{\mu\nu}&=&\tilde{ G_{\mu\nu}}+\nabla_{\sigma}K^{\sigma}_{\;\,\mu\nu}- \nabla_{\nu}K^{\sigma}_{\;\,\mu\sigma}+K^{\sigma}_{\;\,\rho\sigma}K^{\rho}_{\;\,\mu\nu}- K^{\sigma}_{\;\,\mu\rho}K^{\rho}_{\;\,\sigma\nu}\nonumber \\
        &-&\frac{1}{2} g_{\mu\nu}(-2\nabla_{\lambda}K^{\sigma\lambda}_{\;\,\sigma}-K_{\sigma\rho}^{\;\,\rho}K^{\sigma\lambda}_{\;\,\lambda}+K_{\sigma\rho\lambda}K^{\sigma\lambda\rho}).
        \end{eqnarray} 
        In addition, $ E_{\mu\nu} $ can be split into the Riemann part $ \tilde{E}_{\mu\nu} $ and the Cartan part including (con)torsion terms as follows 
        \begin{equation}\label{Ecomp}
          	E_{\mu\nu} = \tilde{E}_{\mu\nu}+Q_{ABCD}n^A 
           	e_{~\mu}^{B}n^C e_{~\nu}^{D}+\frac{1}{3} Q_{AC}e_{~\mu}^{A}e_{~\nu}^{C}-\frac{1}{3}\left( Q_{AC}n^An^C
           	+\frac{1}{4}Q\right)g_{\mu\nu}, 	
         \end{equation}
      where $ Q_{ABCD} $, $ Q_{AC} $, and  $ Q $ are given by  Eqs. $(\ref{torcurv1})$, $(\ref{torcurv2})$ and $ (\ref{torcurv3})$, respectively. Note that the torsion contribution to Eq. (\ref{4dee3}) as expected from the geometric form of  $ G_{\mu\nu} $ (see Eq. $\ref {fdett}$) is included in $E_{\mu\nu} $ as can be seen from (\ref{Ecomp}), whereas $ \Pi_{\mu\nu} $ does not depend on the torsion. This is a decisive result in contrast to \cite{Maier2012}. The contribution of torsion to $ \Pi_{\mu\nu} $ in \cite{Maier2012} is due to the unfortunate result they obtained for the brane extrinsic curvature (see Eq. 28 in \cite{Maier2012}), in contrast to ours (see Eq. \ref{extrintenexp}). \\ 
     On the other hand, the effective Einstein equations (\ref{4dee3}) seems formally the same as those obtained in \cite{da2009braneworld}. The similarity is due to the formal similarity of the junction conditions, and the resulting expression for the extrinsic curvature in terms of the brane energy-momentum tensor leading to the same tensor $ \Pi_{\mu\nu} $. The implicit absence of the torsion term in their junction conditions, however, does not allow to use their approach caution. \\
     Finally let us note that the effective equations (\ref{4dee3}) on the brane are not closed, leading to a continuous exchange of energy and momentum between the brane and the bulk. To take a closer look at this, we note that from (\ref{extrintenexp}) and the Codazzi equation (\ref{Codacci}) it follows that the brane energy-momentum tensor obeys the constraint 
         \begin{equation}\label{divenegbrn}
              \nabla^{\mu}S_{\mu\nu}= \nabla^{\mu}\tau_{\mu\nu}=-2\bar {T}_{AB}n^{A}e^{B}_{~\nu}+\frac{2}{\kappa_5^2}T^{\sigma}_{\;\,\mu\nu}{\cal K}^{\mu}_{\;\,\sigma}.
          \end{equation}
      We therefore see that, in addition to the exchange of energy-momentum between the bulk and the brane, the brane energy-momentum tensor is sourced by the brane torsion. Thus in the presence of torsion, even if there is no energy-momentum transfer from the bulk, i.e. $ \bar {T}_{AB}=0 $, the brane energy-momentum tensor is not conserved.\\
       Next, assuming  $	\bar {T}_{AB}=0 $,  we calculate the divergence of the effective equation (\ref{4dee3}).  Contracting the 4-dimensional Bianchi identities (\ref{torbiaid2}) twice 
        \begin{equation}\label{bian}
                \nabla^{\mu}G_{\mu\nu}=R^{\mu}_{\;\,\sigma}T^{\sigma}_{\;\,\nu\mu}+\frac{1}{2}R^{\mu\sigma}_{~\;\,\rho\nu}T^{\rho}_{\;\,\mu\sigma},
          \end{equation}
       and using (\ref{divenegbrn}),	we find that
         \begin{equation}\label{Edefder}  
           \nabla^{\mu}E_{\mu\nu}=\kappa_5^4 \nabla^{\mu}\Pi_{\mu\nu}-R^{\mu}_{\;\,\sigma}T^{\sigma}_{\;\,\nu\mu}-\frac{1}{2}R^{\mu\sigma}_{~\;\,\rho\nu}T^{\rho}_{\;\,\mu\sigma}+\frac{\kappa_5^2\sigma}{3}T^{\sigma}_{\;\,\mu\nu}{\cal K}^{\mu}_{\;\,\sigma},
          \end{equation}
       indicating that the projected Weyl-Cartan tensor is sourced not only by the quadratic brane energy-momentum terms
       but also by the brane torsion. In other words, in addition to spatial gradients and time derivatives in the
       matter fields, the brane torsion also generates nonlocal gravitational effects in the bulk back-reacting on the brane.

         \section{Conclusion}

   A general standard form of the effective Einstein equations on a 3-brane embedded in an arbitrary 5-dimensional Riemann-Cartan bulk spacetime is the basis of any research in the brane cosmology with torsion. We have looked at different specific approaches in the literature, pinpointing their fallacies or restrictive assumptions. By applying our recently formulated general junction conditions across the brane in the presence of torsion, we have arrived at a standard form for the effective Einstein equation on the $ Z_{2} $ symmetric brane. This form being similar to that already published in the literature, has nevertheless delicate differences in the assumptions and definitions of different terms, as explained in the text.\\

\end{document}